\documentclass[a4paper, reqno, 11pt]{amsart}

\usepackage{amsfonts, amsmath, amssymb, amsthm, amscd, setspace, subfigure,mathtools,stmaryrd}
\usepackage{comment}
\usepackage[heightrounded,textwidth=400pt,top=1in,left=1in,right=1in,bottom=1in]{geometry}
\usepackage{layout}
\usepackage[utf8]{inputenc}
\usepackage{cite}
\usepackage{tikz}
\usepackage{hyperref}
\usetikzlibrary{calc}

\numberwithin{equation}{section}

\newcommand\be{\begin{eqnarray}}
\newcommand\ee{\end{eqnarray}}

\newcommand{\C}{{\mathbb C}}
\newcommand{\R}{{\mathbb R}}

\newcommand{\id}{{\mathbb I}}
\newcommand{\im}{{\rm i\,}}

\theoremstyle{definition}

\author[Krasnov]{Kirill Krasnov}
\email{kirill.krasnov@nottingham.ac.uk, ORCID: 0000-0003-2800-3767}
\address{School of Mathematical Sciences, University of Nottingham, Nottingham, NG7 2RD, UK}

\author[Lipstein]{Arthur Lipstein}
\email{arthur.lipstein@durham.ac.uk}
\address{Department of Mathematical Sciences, Durham University, Durham, DH1 3LE, UK}

\title{Pure connection formalism \\ and \\ Plebanski's second heavenly equation}

\begin{document}

\begin{abstract}\noindent  Plebanski's second heavenly equation reduces the problem of finding a self-dual Einstein metric to solving a non-linear second-order PDE for a single function. Plebanski's original equation is for self-dual metrics obtained as perturbations of the flat metric. Recently, a version of this equation was discovered for self-dual metrics arising as perturbations around a constant curvature background. We provide a new simple derivation of both versions of the Plebanski second heavenly equation. Our derivation relies on the ``pure connection" description of self-dual gravity. Our results also suggest a new interpretation to the kinematic algebra of self-dual Yang-Mills theory, as the Lie algebra of $(0,1)$ vector fields on a $\R^4$ endowed with a complex structure. 
\end{abstract}

\maketitle

\section{Introduction}

Self-dual Yang-Mills (SDYM) and self-dual Einstein gravity (SDG) are very important truncations of full Yang-Mills and gravity, respectively. Both are integrable, in the sense that the most general solution of the non-linear field equations can be written down for example via the perturbiner expansion (see \cite{Rosly:1996vr} for the SDYM case and \cite{Miller:2024oza} for SDGR). The integrability of these theories has also been understood from various other perspectives \cite{Boyer:1985aj,Park:1989vq,Bardeen:1995gk,Dunajski_1998}. Moreover the amplitudes of these theories take a simple form \cite{Bern:1993qk,Bern:1996ja,Bern:1998xc,Bern:1998sv} and certain amplitudes in the full theories can be computed from form factors of the self-dual ones \cite{Costello:2022wso}. Self-dual theories also play a fundamental role twistor theory \cite{Penrose:1976js,Ward:1977ta} and celestial holography \cite{Ball:2021tmb}.

For self-dual theories there exists a field ansatz which reduces the problem of finding solutions of the associated non-linear equations to a PDE on a single function. For SDYM this equation was found in \cite{Parkes:1992rz,Chalmers:1996rq} and for SDGR this is Plebanski's second heavenly equation \cite{Plebanski:1975wn}. These equations make a property known as color/kinematics duality manifest \cite{Monteiro:2011pc} and exhibit an infnite dimensional symmetry \cite{Campiglia:2021srh,Ball:2021tmb}. This can be seen in the existence of the recursion operator which produces solutions of the linearised equations of motion from other such solutions (see for example Appendix C.3 of \cite{Miller:2024oza} for the SDYM case and formulas (3.6)-(3.8) of that reference for the SDGR case). 

Self-dual theories are usually analysed in the light-cone gauge. This is manifest in the treatment in \cite{Parkes:1992rz}, and is effectively what the analysis in \cite{Plebanski:1975wn} amounts to. Light-cone constructions proceed by introducing a set of null coordinates on Minkowski space, two of which are real and two complex. The SDYM ansatz then sets two of the null components of the gauge-field to zero, while the other two are given by partial derivatives of a (Lie-algebra valued) scalar potential, see formula (5) in \cite{Parkes:1992rz}. In the SDGR case one similarly introduces a special null coordinate system, and then a potential for the metric. A drawback of the light-cone approach is that after fixing null coordinates, it becomes difficult to see how the results obtained depend on this choice. See \cite{Campiglia:2021srh,Bonezzi:2023pox} for recent progress in this direction.

A different but related approach to study these theories is the spinor-helicity formalism, see e.g. \cite{Dixon:2013uaa} and references therein, where the polarisation vector of a gluon with 4-momentum $k_\mu$ is expressed as:
\be\label{polarisation}
\varepsilon^+_{MM'} = \frac{ q_M k_{M'}}{\langle qk\rangle},
\ee
where $A,A'$ are 2-component spinor indices of opposite chirality, $q_M$ is a 2-component reference spinor, $\langle qk\rangle = q^M k_M$, and $k_M, k_{M'}$ are the momentum spinors arising from the fact that the momentum $k_{MM'}$ is null and therefore $k_{MM'}= k_M k_{M'}$. A similar representation exists for graviton polarisations, which are essentially tensor products of gluon polarisations. Note that the reference spinor encodes gauge symmetry and can be chosen arbitrarily. Light-cone gauge then corresponds to (\ref{polarisation}) for a specific choice of the reference spinor, as is manifest in e.g. formulas following (2.13) in \cite{Miller:2024oza}. In this paper, we will use a covariant light-cone ansatz where the self-dual gauge field is of the form
\be\label{A-ansatz-intro}
A_\mu = \bar{\Omega}_{\mu}{}^{\nu} \partial_\nu \phi,
\ee
where $\bar{\Omega}_{\mu\nu}$ is a decomposable (complex) self-dual 2-form and $\phi$ is an arbitrary ``potential" function. In terms of spinors the 2-form $\bar{\Omega}_{\mu\nu}$ is of the form $\bar{\Omega}_{MM'NN'} = q_M q_N \epsilon_{M'N'}$, where $\epsilon_{M'N'}$ is the spinor metric (not to be confused with the polarisation vector $\varepsilon^+_{MM'}$). In momentum space, the ansatz becomes
\be
A_{MM'} = q_M q^N \epsilon_{M'}{}^{N'} k_{N} k_{N'} \phi = q_M k_{M'} \langle q k\rangle \phi,
\ee
which coincides with (\ref{polarisation}) for a suitable choice of $\phi$. The ansatz (\ref{A-ansatz-intro}) therefore combines the benefits of both the light-cone and spinor-helicity formalisms. It also has a nice geometric interpretation: as we will see below, after analytically continuing to Euclidean signature, the choice of $\bar{\Omega}$ encodes a choice of complex structure in $\R^4$. Such an approach is far from new. Its version in classical electrodynamics, where the electromagnetic field is parametrised by a derivative of a scalar, is known as the Hertz and Debye potentials. These techniques have also been applied to gravity since the 70's, see e.g. \cite{Kegeles:1979an} and references therein. A similar approach for self-dual theories was also recently developed in \cite{Campiglia:2021srh}.

Using \eqref{A-ansatz-intro}, we show that the SDYM equation reduces to (\ref{sd-eqs-YM-flat}). This is equivalent to results found long ago except that our choice of null coordinates is kept arbitrary and parameterised by the tensor $\bar{\Omega}$. As a bonus, this formalism suggests a new mathematical interpretation of the kinematic algebra of SDYM theory. This algebra was first described in \cite{Monteiro:2011pc}. Our approach gives a spacetime covariant interpretation of this algebra. We then apply the same ideas to SDGR using the pure connection formalism \cite{Krasnov:2016emc,Krasnov:2021cva} and re-derive Plebanski's second heavenly equation in the form (\ref{sd-eqs-flat}). Finally, we consider the problem of self-dual gravity in a constant curvature background and re-derive the scalar equation of motion first obtained in \cite{Lipstein:2023pih}. A striking feature of our SDG calculation is its relative simplicity. Indeed, the scalar equation of motion for SDG in constant curvature background was previously obtained from a very tedious calculation, but using the formalism developed in this paper it can be derived in only a few lines. This suggests that the pure connection formalism may provide a promising avenue for computing observables in (Anti) de Sitter space ((A)dS), which are crucial for the study holography and cosmology. Due to their simplicity, self-dual theories provide the ideal setting for developing new calculational techniques in these backgrounds. In contrast to flat space however, very little is known about the observables of self-dual theories in (A)dS, although the first steps in computing their boundary correlation functions were recently taken in \cite{Jain:2024bza,Chowdhury:2024dcy}. It was also recently shown that SDG in AdS$_4$ exhibits color/kinematics duality and enjoys a $w_{1+\infty}$ symmetry \cite{Lipstein:2023pih,Bittleston:2024rqe,Taylor:2023ajd}. For related recent work, see \cite{Neiman:2023bkq}.

\section{Self-dual YM and GR in flat space}

\subsection{SDYM}

As is well-known, the problem of finding self-dual YM fields can be reduced to a PDE for a single Lie-algebra-valued potential function. We start by recalling a derivation of this PDE, to establish notations but also to point out some facts that do not seem to have been noticed before. 

Self-dual, or rather anti-self-dual gauge fields, which we call self-dual for reasons of language economy, are solutions to the following set of first-order PDE's on the gauge field:
\be\label{self-duality-YM}
(F^a_{\mu\nu})^+ =0.
\ee
Here 
\be
F^a_{\mu\nu} = \partial_{\mu} A^a_\nu - \partial_\nu A^a_\mu + f^{abc} A^b_\mu A^c_\nu
\ee
is the field strength, and the plus superscript denotes the projection onto the self-dual part. 

The projection onto the self-dual part can be computed as follows. We introduce a basis in the space of self-dual 2-forms. We will work in Euclidean signature, which reduces the number of factors of the imaginary unit appearing in the calculations. In flat space, the basis of self-dual 2-forms is given by
\be
\Sigma^i = dt \wedge dx^i - \frac{1}{2} \epsilon^{ijk} dx^j \wedge dx^k.
\ee
These objects satisfy the following set of relations:
\be
\Sigma^i \wedge \Sigma^j \sim \delta^{ij}.
\ee
Also, if one raises one of the indices of $\Sigma^i_{\mu\nu}$ to form an object $\Sigma^i_\mu{}^\nu$, this can be viewed as an endomorphism of the (co-)tangent space. The triple of arising endomorphisms satisfies the algebra of the imaginary quaternions
\be\label{algebra-Sigma}
\Sigma^i{}_\mu{}^\alpha \Sigma^j_\alpha{}^\nu = - \delta^{ij} \delta_\mu{}^\nu + \epsilon^{ijk} \Sigma^k_\mu{}^\nu.
\ee

Using the basis of self-dual 2-forms introduced above, the self-duality equations (\ref{self-duality-YM}) can be rewritten as
\be\label{self-duality-eqs-YM}
\Sigma^{i\mu\nu} F^a_{\mu\nu}=0.
\ee

\subsection{Complex self-dual 2-forms and a complex structure}

For what follows, it will be very convenient to introduce the following complex linear combinations of the self-dual 2-forms:
\be\label{complex-Omega-omega}
\Omega:= \Sigma^1+\im \Sigma^2, \quad \bar{\Omega} =  \Sigma^1-\im \Sigma^2, \quad \omega=\Sigma^3.
\ee
We note that $\Sigma^{1,2,3}$ are only real in Euclidean signature, and only in this signature is $\bar{\Omega}$ the complex conjugate of $\Omega$. The new complex self-dual 2-forms satisfy
\be\label{algebra-Omega-omega}
\Omega\wedge \Omega=0, \qquad \Omega\wedge \omega=0, \qquad \Omega\wedge \bar{\Omega} = 2 \omega^2.
\ee
These equations have an interpretation as defining a complex structure on $\R^4$. Indeed, the first equation states that the 2-form $\Omega$ is decomposable. It thus defines a 2-dimensional subspace in the space of 1-forms $\Lambda^1$. One can declare this subspace to be the subspace $\Lambda^{(1,0)}$ of $(1,0)$ forms of some complex structure $J:\R^4\to \R^4, J^2=-\id$. This defines the complex structure. The second equation in (\ref{algebra-Omega-omega}) then says that $\omega\in \Lambda^{(1,1)}\subset \Lambda^2$. Raising one of the indices of $\omega$ with the metric and using the algebra (\ref{algebra-Sigma}) gives $\omega_\mu{}^\alpha \omega_\alpha{}^\nu = - \delta_\mu{}^\nu$, which identifies $\omega_\mu{}^\nu$ as the complex structure. It is thus clear that the choice we have made in constructing (\ref{complex-Omega-omega}) is the choice of a complex structure on $\R^4$. 

Using the complex self-dual 2-forms, the self-duality equations (\ref{self-duality-eqs-YM}) get rewritten as
\be\label{self-duality-YM-complex}
\Omega^{\mu\nu} F^a_{\mu\nu}=0, \qquad \bar{\Omega}^{\mu\nu} F^a_{\mu\nu}=0, \qquad \omega^{\mu\nu} F^a_{\mu\nu}=0.
\ee

\subsection{Potential for self-dual gauge fields}

We now reduce the problem of constructing solutions to (\ref{self-duality-YM-complex}) to the problem of solving a PDE for a single function. To this end, we parametrise the gauge field  as
\be\label{param-A-YM}
A^a_\mu = \bar{\Omega}_\mu{}^\nu \partial_\nu \phi^a. 
\ee
Writing this equation in components shows that it is exactly the same ansatz for the gauge field that is made when analysing the SDYM in the light-cone gauge \cite{Parkes:1992rz}. One can refer to the ansatz (\ref{param-A-YM}) as a covariant light-cone gauge, because the choice of the light-cone coordinates is now explicitly parametrised by the tensor $\bar{\Omega}_\mu{}^\nu$. One more novelty is that we work in Euclidean signature where this tensor has a clear geometric interpretation of defining a complex structure. 

The ansatz (\ref{param-A-YM}) automatically satisfies two of the three equations in (\ref{self-duality-YM-complex}). Indeed, we have the following algebraic properties of the complex 2-forms
\be\label{algebra-omega-Omega}
\bar{\Omega}_\mu{}^\alpha \bar{\Omega}_\alpha{}^\nu =0, \qquad \bar{\Omega}_\mu{}^\alpha \omega_\alpha{}^\nu = -\im \bar{\Omega}_\mu{}^\nu. 
\ee
The first of these shows that $\bar{\Omega}^{\mu\nu} F^a_{\mu\nu}=0$ is automatically satisfied, while using the second $\omega^{\mu\nu} F^a_{\mu\nu}$ becomes
\be\nonumber
\omega^{\mu\nu} ( 2 \partial_\mu \bar{\Omega}_\nu{}^\alpha \partial_\alpha \phi^a + f^{abc} \bar{\Omega}_\mu{}^\alpha \partial_\alpha \phi^b \bar{\Omega}_\nu{}^\beta \partial_\beta \phi^b) 
= \im ( - 2 \bar{\Omega}^{\mu\alpha} \partial_\mu \partial_\alpha \phi^a + f^{abc} \bar{\Omega}^{\nu\alpha} \partial_\alpha \phi^b \bar{\Omega}_\nu{}^\beta \partial_\beta \phi^b) =0,
\ee
where we again used the first relation in (\ref{algebra-Omega-omega}). 

The self-duality equation then reduces to the condition $\Omega^{\mu\nu} F^a_{\mu\nu}=0$, or, in terms of parametrisation (\ref{param-A-YM})
\be
\Omega^{\mu\nu} ( \partial_\mu \bar{\Omega}_\nu{}^\alpha \partial_\alpha \phi^a + f^{abc} \bar{\Omega}_\mu{}^\alpha \partial_\alpha \phi^b \bar{\Omega}_\nu{}^\beta \partial_\beta \phi^c)=0.
\ee
Using the algebra of 2-forms 
\be\label{Omega-bar-Omega}
\Omega^{\mu\alpha}  \bar{\Omega}_\alpha{}^\nu = - 2 g^{\mu\nu} - 2\im \omega^{\mu\nu},
\ee
this becomes
\be\label{sd-eqs-YM-flat}
\Box \phi^a =\frac{1}{2} f^{abc} \bar{\Omega}^{\mu\nu} \partial_\mu \phi^b \partial_\nu \phi^c.
\ee
This is the well-known PDE for the function $\phi^a$, see e.g. \cite{Parkes:1992rz} for essentially the same derivation as the one above, except that we parametrise different possible choices of the light-cone gauge by the choice of a complex structure, and in particular, tensors $\Omega_{\mu\nu}, \bar{\Omega}_{\mu\nu}, \omega_{\mu\nu}$. 

\subsection{New interpretation of the kinematic algebra of the SDYM}

We note that the parametrisation (\ref{param-A-YM}) provides a new interpretation to the kinematic algebra of self-dual YM theory. In \cite{Monteiro:2011pc} this algebra is interpreted as that of area-preserving diffeomorphisms of a certain two-dimensional space. A simple calculation related to (\ref{param-A-YM}) provides a somewhat different interpretation. 

Consider vector fields on $\R^4$ that are of the form
\be
X_\phi^\mu = \bar{\Omega}^{\mu\nu} \partial_\nu \phi.
\ee
These can be interpreted as Hamiltonian vector fields for the Poisson structure determined by the Poisson bivector $\bar{\Omega}^{\mu\nu}$. Such vector fields are necessarily $(0,1)$ vector fields, and so vector fields of this type can be referred to as Hamiltonian $(0,1)$ vector fields. A standard calculation shows that Hamiltonian vector fields form a Lie subalgebra of the algebra of all vector fields. Indeed, we have
\be
[X_{\phi_1},X_{\phi_2}] = \bar{\Omega}^{\alpha\beta} \partial_\beta \phi_1 \partial_\alpha ( \bar{\Omega}^{\mu\nu} \partial_\nu \phi_2) - \bar{\Omega}^{\alpha\beta} \partial_\beta \phi_2 \partial_\alpha ( \bar{\Omega}^{\mu\nu} \partial_\nu \phi_1)
\ee
Here we are assuming all background tensors (such as $\bar{\Omega}^{\alpha\beta}$) to be constant, and so can be pulled out from under the derivative, so we have
\be\nonumber
[X_{\phi_1},X_{\phi_2}] = \bar{\Omega}^{\alpha\beta} \bar{\Omega}^{\mu\nu} ( \partial_\beta \phi_1 \partial_\alpha  \partial_\nu \phi_2 -\partial_\beta \phi_2 \partial_\alpha \partial_\nu \phi_1).
\ee
We can rewrite this as
\be\nonumber
[X_{\phi_1},X_{\phi_2}] = \bar{\Omega}^{\alpha\beta} \bar{\Omega}^{\mu\nu} ( \partial_\beta \phi_1 \partial_\alpha  \partial_\nu \phi_2 +\partial_\beta \partial_\nu \phi_1 \partial_\alpha \phi_2 ) = \bar{\Omega}^{\alpha\beta} \bar{\Omega}^{\mu\nu}  \partial_\nu ( \partial_\beta \phi_1 \partial_\alpha  \phi_2).
\ee
Taking $ \bar{\Omega}^{\alpha\beta}$ back under the derivative sign we can rewrite this as
\be
[X_{\phi_1},X_{\phi_2}] = X_{[\phi_1,\phi_2]}, \qquad [\phi_1,\phi_2] := \bar{\Omega}^{\alpha\beta} ( \partial_\beta \phi_1 \partial_\alpha  \phi_2 ).
\ee
This is of course a version of the standard calculation in Poisson geometry, further simplified by the fact that we have assumed the bivector $\bar{\Omega}^{\mu\nu}$ to be constant. 

The Lie algebra with the bracket $[\phi_1,\phi_2]$ is precisely the kinematic Lie algebra of SDYM theory. We have thus exhibited a homomorphism from this Lie algebra into the Lie algebra of vector fields on $\R^4$, as the subalgebra of Hamiltonian vector fields for the Poisson structure with $\bar{\Omega}^{\mu\nu}$ as the Poisson bivector. Alternatively, because such Hamiltonian vector fields span all of $(0,1)$ vector fields, we can say that the kinematic Lie algebra of SDYM is the Lie algebra of all $(0,1)$ vector fields on $\R^4$ endowed with a complex structure. This interpretation appears to be new. 

\subsection{Flat space SDGR in the complex basis}

We now consider flat self-dual GR, in the formulation described in \cite{Krasnov:2021cva}. The calculation in this subsection is a pre-cursor to a more involved calculation in constant curvature space in the next section. Viewing flat SD metrics as perturbations around flat space, the SDGR action takes the following form:
\be
S[A,\Psi] = \int \Psi^{ij}(\Sigma^i + dA^i)(\Sigma^j+ dA^j).
\ee
Here $\Psi^{ij}$ is a $3\times 3$ symmetric tracefree matrix, $\Sigma^i$ is a triple of self-dual 2-forms for the background (flat) metric, and $A^i$ is a triple of 1-forms. The non-linear equation for self-dual gravitons is obtained by varying the above action with respect to $\Psi^{ij}$, and is given by
\be\label{flat-SDGR-eqs}
(\Sigma^i + dA^i)(\Sigma^j+ dA^j)\sim \delta^{ij}.
\ee
The metric is then obtained from the triple of 2-forms $B^i:= \Sigma^i + dA^i$ using the Urbantke formula \cite{Urbantke:1984eb}
\be\label{Urbantke}
g(\xi,\eta) v_g = \frac{1}{6} \epsilon^{ijk} i_\xi B^i \wedge i_\eta B^j \wedge B^k.
\ee
Here $g(\xi,\eta)$ is the metric pairing of two vector fields $\xi,\eta\in TM$, and $v_g$ is the volume form for the metric. Both sides are top forms, and the right-hand side defines a symmetric expression in $(\xi,\eta)$, which is identified with the metric pairing. 

As in the case of SDYM, it will be beneficial to rewrite this action in a different basis, introducing complex linear combinations of all the fields. 
We introduce the already familiar complex self-dual 2-forms $\Omega,\bar{\Omega},\omega$, as well as complex 1-forms
\be
A: = A^1 + \im A^2, \quad \bar{A} := A^1-\im A^2, \qquad A^3:=a. 
\ee
In this complex basis, the flat SDGR field equations (\ref{flat-SDGR-eqs}) become the following system of equations:
\be
(\Omega + dA)\wedge (\Omega+dA)=0, \qquad (\Omega + dA)\wedge (\omega+da)=0, \\ \nonumber (\bar{\Omega} + d\bar{A})\wedge (\Omega+dA)= 2(\omega+da)^2,
\ee
together with their complex conjugates. 

\subsection{Plebanski's second heavenly equation}

We make the following ansatz for all the fields
\be\label{a-flat}
a=\bar{A}=0, \qquad A_\mu =  \bar{\Omega}_\mu{}^\nu \partial_\nu \phi.
\ee
Note that this is completely analogous to the ansatz (\ref{param-A-YM}) in the SDYM case. 
Then there are three equations to satisfy:
\be\label{eqs-flat}
(\Omega+dA)\wedge (\Omega+ dA)=0, \qquad dA \wedge \omega = 0, \qquad dA \wedge \bar{\Omega}=0.
\ee
The second and third are satisfied because of the identities satisfied by the 2-forms. Indeed, using the self-duality of $\omega,\Omega$ we can rewrite the second equation as
\be
\omega^{\mu\nu} \partial_\mu \bar{\Omega}_\nu{}^\alpha \partial_\alpha \phi =0.
\ee
The action of $\omega_\mu{}^\nu$ on any $(0,1)$ form corresponds to multiplying it by $\im$. This gives
\be
\omega^{\mu\nu} \partial_\mu \bar{\Omega}_\nu{}^\alpha \partial_\alpha \phi = \im \bar{\Omega}^{\mu\alpha} \partial_\mu \partial_\alpha \phi =0,
\ee
because partial derivatives commute and $\bar{\Omega}^{\mu\nu}$ is anti-symmetric. 
For the third equation we have
\be
\bar{\Omega}^{\mu\nu} \partial_\mu \bar{\Omega}_\nu{}^\alpha \partial_\alpha \phi =0.
\ee
It is satisfied because two copies of $\bar{\Omega}$ contracting in a pair of indices gives zero, see (\ref{algebra-omega-Omega}). 

The only non-trivial equation in (\ref{eqs-flat}) is then the first one, which can be re-written as
\be
2\Omega dA + dA dA =0.
\ee
Rewriting this in index notation gives
\be
2\Omega^{\mu\nu}  \partial_\mu \bar{\Omega}_\nu{}^\alpha \partial_\alpha \phi  +  \epsilon^{\mu\nu\rho\sigma} \partial_\mu \bar{\Omega}_\nu{}^\alpha \partial_\alpha \phi 
\partial_\rho \bar{\Omega}_\sigma{}^\beta \partial_\beta \phi =0.
\ee
Using (\ref{Omega-bar-Omega}) we see that we get the box operator on the left-hand side. To simplify the right-hand side we will use the self-duality of $\bar{\Omega}$, which implies that 
\be\label{epsilon-omega}
\epsilon^{\mu\nu\rho\sigma} \bar{\Omega}_{\mu\alpha} = \delta_\alpha{}^{\nu} \bar{\Omega}^{\rho\sigma} + \delta_\alpha{}^{\rho} \bar{\Omega}^{\sigma\nu}+\delta_\alpha{}^{\sigma} \bar{\Omega}^{\nu\rho}.
\ee
Applying this identity to the contraction of $\epsilon$ with e.g. the first copy of $\bar{\Omega}$ gives three terms, two of which will give a contraction of two copies of $\bar{\Omega}$, which vanishes. This leaves the following equation:
\be\label{sd-eqs-flat}
\Box \phi = \frac{1}{4} \bar{\Omega}^{\mu\rho} \bar{\Omega}^{\alpha\beta} \partial_\mu \partial_\alpha \phi \partial_\rho \partial_\beta \phi.
\ee
This is Plebanski's second heavenly equation. The derivation we presented is new, and is significantly simpler than the one available in the literature, see e.g. Appendix B of \cite{Miller:2024oza}. Note that the double copy structure of this equation as compared to (\ref{sd-eqs-YM-flat}) is manifest. In the next section we obtain an appropriate non-flat covariant version of this equation for SDGR in de Sitter space. 

\section{Self-dual gravity in hyperbolic space}

Our task now is to obtain a version of (\ref{sd-eqs-flat}) for SDGR in a constant curvature background. We will use a pure connection formalism, which allows one to describe 4d Einstein gravity with non-zero cosmological constant $\Lambda$  (and its self-dual truncation) as a particular theory of an ${\rm SU}(2)$ gauge field. We start by reviewing this formalism. 

\subsection{Chiral pure connection description of GR and SDGR}

Let us start with some generalities. General Relativity is normally described by a formalism in which a field encoding a metric (this can be the metric field itself, or e.g. the frame field) is subject to second-order PDE's, notably the Einstein equations. There exists, however, a first order formalism, in which the connection field that defines the appropriate covariant derivative is treated as an independent one. An action is then written that depends both on the metric and connection fields. In the case of the metric description this is called the Palatini formalism, in which the metric appears in the action together with the field $\Gamma_{\mu\nu}{}^\rho$ which gets identified with the Christoffel symbol on-shell. In the case of the tetrad description one has the Einstein-Cartan action, in which the tetrad field appears in the action together with the spin connection. See e.g. \cite{Krasnov:2020lku}, sections 2.4 and 3.2 for more details. One can integrate out the connection field from these Lagrangians by solving the (algebraic) equations of motion for this field and substituting the result back into the Lagrangian. One then recovers the usual second-order in derivatives formalism. However, when $\Lambda\not=0$, there is an alternative. One can instead solve an (algebraic) field equation obtained by varying the action with respect to the metric, and substitute the result back into the Lagrangian. One then obtains a ``pure connection" description of gravity. This trick is only possible when $\Lambda\not=0$, which is reflected by the fact that one obtains a factor of $1/\Lambda$ in front of the resulting pure connection Lagrangian. In the metric description of gravity, the corresponding pure connection formalism is that of Eddington description of GR, see section 2.5 of \cite{Krasnov:2020lku}. In the case of the frame formalism, the corresponding pure connection formulation is described in section 3.4 of that reference. 

In addition to the metric and the frame formalisms, in 4D there is also the Plebanski formalism that uses a triple of 2-forms to encode the metric, see section 5 of  \cite{Krasnov:2020lku}. This is a first-order description of gravity, in which the action depends on both a collection of 2-forms $\Sigma^i$, as well as an independent collection of 1-forms $A^i$. When $\Lambda\not=0$, the 2-form field $\Sigma^i$ can be integrated out from the action, resulting in a pure connection description of gravity. The most useful version of this description is one where a certain Lagrange multiplier field is also kept. The GR action then takes the form
\be\label{GR-pure-con}
S[A,\Psi] = \frac{1}{16\pi G} \int \left( \Psi^{ij} + \frac{\Lambda}{3} \delta^{ij}\right)^{-1} F^i \wedge F^j.
\ee
Here $\Psi^{ij}$ is a symmetric, trace-free Lagrange multiplier field, $A^i$ is an ${\rm SO}(3)$ connection, and $F^i$ is its curvature 2-form. 
The Lagrange multiplier field can also be integrated out, producing an action first described in \cite{Krasnov:2011pp}. But the action in the way we wrote it will be more convenient for the purposes of this paper. 

One can expand the tensor $\left( \Psi^{ij} + \frac{\Lambda}{3} \delta^{ij}\right)^{-1}$ appearing in the action in powers of $\Psi^{ij}$. The first term that results is the topological $F^i\wedge F^i$ term. The second term gives the action
\be\label{SDGR-Lagr}
S_{\rm SDGR}[A,\Psi] = \int \Psi^{ij} F^i \wedge F^j.
\ee
It can then be shown that the critical points of this action are Einstein (with non-zero $\Lambda$) half Weyl-flat manifold, i.e. manifolds where the Ricci curvature satisfies $R_{\mu\nu}=\Lambda g_{\mu\nu}$ and half of the Weyl curvature vanishes $W^+=0$. More precisely, the statement is that any ${\rm SO}(3)$ connection that satisfies the set of first-order PDE's 
\be
F^i\wedge F^j\sim \delta^{ij}
\ee
defines a metric that is Einstein (with non-zero $\Lambda$) and half Weyl-flat. This metric is defined by introducing a triple of 2-forms
\be
B^i : = \frac{1}{\Lambda} F^i,
\ee
and substituting this into Urbantke formula (\ref{Urbantke}). For details of the proof of this statement, the reader is advised to consult e.g. section 6.3 of \cite{Krasnov:2011pp}. 

The pure connection description of $\Lambda\not=0$ SDGR makes it clear that this theory is a truncation of full GR in which one keeps only the first non-trivial term in the expansion of the full GR action (\ref{GR-pure-con}) into powers of $\Psi^{ij}$. We thus get a very clear description of how SDGR sits inside full GR. 

Our task now is to describe self-dual gravitons as certain perturbations of the connection $A^i$ around the constant curvature background. We start by describing the background.

\subsection{Hyperbolic space}

We will use the conformally flat description of the hyperbolic space, for which the metric is
\be
ds^2 = \frac{1}{t^2}( dt^2 + dx^2 + |dz|^2).
\ee
Here $t,x$ are real coordinates, and $z\in \C$ is a complex coordinate on $\R^2\sim \C$. We use the Plebanski formalism, in which the metric is encoded into a triple of 2-forms $\Sigma^i, i=1,2,3$, with 
\be
\Sigma^i = e^{4}\wedge e^i - \frac{1}{2} \epsilon^{ijk} e^j\wedge e^k,
\ee
where $e^{4,1,2,3}$ are the frame 1-forms. In Plebanski formalism one then proceeds to find the connection 1-forms $A^i$ from the equations $d\Sigma^i + \epsilon^{ijk} A^j\wedge \Sigma^k=0$. 

An equivalentapproach that is more useful for the problem at hand is to use a complex decomposable 2-form, as well as a real 2-form $\omega: = \Sigma^3, \Omega:= \Sigma^1+\im \Sigma^2$. For the hyperbolic space metric these are given by
\be
\omega= \frac{1}{t^2}( dt dx + \frac{1}{2\im} dz\wedge d\bar{z}), \qquad \Omega = \frac{1}{t^2}( dt - \im dx) dz.
\ee
The corresponding connection is encoded by a real 1-form $A^3$, and a complex 1-form $A$. They are determined by the following equations:
\be\label{compat}
d\omega + \frac{1}{2\im}(\bar{A}\Omega - A\bar{\Omega}) =0, \qquad d\Omega - \im A\omega +\im A^3 \Omega =0,
\ee
and are given by
\be
A= \frac{dz}{t}, \qquad A^3=\frac{dx}{t}.
\ee
The curvature 2-forms are
\be\label{curvatures}
F= dA+ \im A^3 A, \qquad \bar{F} = d\bar{A} - \im A^3 \bar{A}, \qquad F^3= dA^3 - \frac{1}{2\im} A \bar{A}.
\ee
A simple computation gives 
\be\label{backgr}
F= -\Omega, \qquad F^3=-\omega,
\ee
which confirms that this is a space of negative scalar curvature. 

\subsection{$\Lambda\not=0$ SDGR in the pure connection formalism}

As we have already described, SDGR with $\Lambda\not=0$ is described in the pure connection formalism by a triple of connection 1-forms $A^i$. The SDGR field equations are then $F^i\wedge F^j\sim\delta^{ij}$. However, it will be more convenient to switch to the complex basis, in which we instead consider a complex 1-form $A$ and a real 1-form $A^3$. The curvatures are given by (\ref{curvatures}), and the SDGR equations take the form
\be\label{eqs}
F\wedge F=0, \qquad F\wedge F^3=0, \qquad \bar{F}\wedge \bar{F}=0, \qquad \bar{F}\wedge F^3=0, \qquad F\wedge \bar{F} = 2 F^3\wedge F^3.
\ee 

\subsection{SDGR in a space of constant curvature}

We now consider the problem of describing SDGR with $\Lambda\not=0$. We consider the connection configuration given by 
\be
A= \frac{dz}{t}+a, \qquad A^3= \frac{dx}{t}, \qquad \bar{A} = \frac{d\bar{z}}{t},
\ee
where
\be\label{a-Omega}
a_\mu = \frac{1}{2t} \bar{\Omega}_\mu{}^\nu \partial_\nu \phi,
\ee
or explicitly 
\be\label{a}
a= \frac{1}{t} \partial_z \phi ( dt+ \im dx) - \frac{1}{2t}( \partial_t \phi + \im \partial_x \phi) d\bar{z},
\ee
and $\phi$ is a field that depends on all 4 coordinates $t,x, z,\bar{z}$. Note that this is a chiral configuration in the sense that $\bar{A}\not = A^*$. Note that our connection ansatz is
completely analogous to (\ref{a-flat}), with the only difference being the presence of the factor of $1/t$ in front. The need for this factor is not surprising, and reflects the fact that the background is curved. 

For completeness, let us spell out how (\ref{a}) is obtained from (\ref{a-Omega}). We have
\be
\bar{\Omega} = \frac{1}{t^2} (dt+\im dx) \wedge d\bar{z}.
\ee
Raising one of its indices with the inverse metric 
\be
g^{\mu\nu} = t^2 \left( \left(\frac{\partial}{\partial t}\right)^\mu \left(\frac{\partial}{\partial t}\right)^\nu +  \left(\frac{\partial}{\partial x}\right)^\mu \left(\frac{\partial}{\partial x}\right)^\nu
+ 2  \left(\frac{\partial}{\partial z}\right)^\mu \left(\frac{\partial}{\partial \bar{z}}\right)^\nu+ 2  \left(\frac{\partial}{\partial \bar{z}}\right)^\mu \left(\frac{\partial}{\partial z}\right)^\nu\right),
\ee
gives the following object:
\be\label{bar-Omega-ud}
\bar{\Omega}_\mu{}^\nu = 2(dt+\im dx)_\mu \left(\frac{\partial}{\partial z}\right)^\nu - d\bar{z}_\mu \left( \left(\frac{\partial}{\partial t}\right)^\nu + \im \left(\frac{\partial}{\partial x}\right)^\nu \right).
\ee
It is now immediate to see that (\ref{a}) is indeed given by (\ref{a-Omega}). 

Using (\ref{backgr}), the curvature components take the following form
\be\label{new-curvatures}
F= -\Omega + da+ \im A^3 \wedge a, \qquad \bar{F}=-\bar{\Omega}, \qquad F^3= -\omega - \frac{1}{2\im} a\wedge \bar{A}.
\ee
A computation shows that all of the equations in (\ref{eqs}) are satisfied by this connection ansatz, apart from the equation $F\wedge F=0$, which takes the following form:
\be\label{arthur-formula}
\frac{1}{t}(4\partial_z\partial_{\bar{z}}\phi + \partial_x^2 \phi + \partial_t^2 \phi) = \partial^2_z \phi ( \partial_u - \frac{2}{t}) \partial_u \phi - (\partial_u \partial_z\phi)(\partial_u - \frac{2}{t}) \partial_z \phi,
\ee
where
\be
\partial_u = \partial_t +\im \partial_x.
\ee
This is the same as the equations (34), (36) in \cite{Lipstein:2023pih}. A similar formula written in terms of spinors also appears in \cite{Neiman:2023bkq} (see equation (54) of that reference). The last reference also used the pure connection formalism. The novelty here is that we do not need spinors in our treatment. The calculation that verifies all the equations in (\ref{eqs}), and also obtains a covariant version of the equation (\ref{arthur-formula}) is spelled out below. 

\subsection{Computation of $d a + \im A^3 a$}

We now want to perform the calculation leading to (\ref{arthur-formula}) by hand, and also rewrite this formula in a way analogous to (\ref{sd-eqs-flat}). 
The new curvature 2-forms are given by (\ref{new-curvatures}). This means that $\bar{F}\bar{F}=0$ is unchanged and $\bar{F}F^3 \sim \bar{\Omega}\bar{A}=0$ continues to hold because both $\bar{\Omega},\bar{A}$ contain $d\bar{z}$. The other equations need some work to be established. To this end, we first compute the object $d a + \im A^3 a$.

We have
\be
d a + \im A^3 a = d \left( \frac{1}{t} \partial_z \phi ( dt+ \im dx) - \frac{1}{2t}( \partial_t \phi + \im \partial_x \phi) d\bar{z}\right) \\ \nonumber
+ \frac{\im}{t^2} dx  \left( \partial_z \phi ( dt+ \im dx) - \frac{1}{2}( \partial_t \phi + \im \partial_x \phi) d\bar{z}\right).
\ee 
This can be written as
\be
 \frac{1}{2t} \bar{\Omega}_{[\nu}{}^\alpha \partial_{\mu]} \partial_\alpha \phi- \frac{1}{t^2} (dt-\im dx)_{[\mu} 
\left( ( dt+ \im dx)_{\nu]} \partial_z \phi - \frac{1}{2}d\bar{z}_{\nu]} ( \partial_t + \im \partial_x) \phi \right) \\ \nonumber
= \frac{1}{2t} \bar{\Omega}_{[\nu}{}^\alpha \partial_{\mu]} \partial_\alpha \phi - \frac{1}{2t^2} (dt-\im dx)_{[\mu} \bar{\Omega}_{\nu]}{}^\alpha \partial_\alpha \phi.
\ee
We can now easily compute
\be\label{bar-Omega-da}
\bar{\Omega}^{\mu\nu}( \partial_{\mu} a_{\nu} + \im A^3_{\mu} a_{\nu} ) =0, \\ \label{omega-da}
\omega^{\mu\nu}( \partial_{\mu} a_{\nu} + \im A^3_{\mu} a_{\nu} ) = -2\im  \partial_z\phi, 
\\ \label{Omega-da}
\Omega^{\mu\nu}( \partial_{\mu} a_{\nu} + \im A^3_{\mu} a_{\nu} ) = \frac{1}{2t} \Omega^{\mu\nu}  \bar{\Omega}_{\nu}{}^\alpha \partial_{\mu} \partial_\alpha \phi = - t \Box\phi,
\ee
where we used the algebra of 2-forms to get the last equality in the last line. The box here is the flat metric box. The contraction of the first derivative terms with both $\Omega^{\mu\nu}$ and $\bar{\Omega}^{\mu\nu}$ vanishes.

\subsection{Analysis of equations continued}

We can use the self-duality of $\omega, \bar{\Omega}$ to rewrite the equation $2F^3 F^3 = F\bar{F}$ as
\be\label{eq-1}
  \frac{2}{\im} \omega^{\mu\nu} a_\mu \bar{A}_\nu = - \bar{\Omega}^{\mu\nu} ( \partial_\mu a_\nu + \im A^3_\mu a_\nu).
  \ee
  However, both $a_\mu$ and $ \bar{A}_\mu$ are $(0,1)$ forms, and the left-hand side is their contraction, which vanishes. The right-hand side vanishes, as we confirmed in (\ref{bar-Omega-da}). This verifies that the equation $2F^3 F^3 = F\bar{F}$ is satisfied. 

Let us now analyse the equation $FF^3=0$. Using self-duality we can rewrite this as
\be\label{eq-2}
\frac{1}{2\im} \Omega^{\mu\nu} a_\mu \bar{A}_\nu = \omega^{\mu\nu} ( \partial_\mu a_\nu + \im A^3_\mu a_\nu).
\ee
The right-hand side of this was already computed in (\ref{omega-da}). 
The left-hand side is equal to
\be
\frac{1}{4\im t} \Omega^{\mu\nu} \bar{\Omega}_\mu{}^\alpha \partial_\alpha \phi \bar{A}_\nu=\frac{1}{2\im t} ( g^{\nu\alpha} + \im \omega^{\nu\alpha}) \bar{A}_\nu \partial_\alpha \phi =
\frac{1}{\im t}  g^{\nu\alpha} \bar{A}_\nu \partial_\alpha \phi = -2\im \partial_z\phi,
\ee
where we used the algebra of 2-forms to get the first equality and the fact that $\omega$ acts on $(0,1)$ forms by multiplication with $\im$ to get the second. Thus, the equation $FF^3=0$ is satisfied. 

We now need to analyse the last equation $F\wedge F=0$, which becomes
\be
2 \Omega^{\mu\nu} ( \partial_\mu a_\nu + \im A^3_\mu a_\nu) = \epsilon^{\mu\nu\rho\sigma} ( \partial_\mu a_\nu + \im A^3_\mu a_\nu) ( \partial_\rho a_\sigma + \im A^3_\rho a_\sigma).
\ee 
The left-hand side was already computed in (\ref{Omega-da}). For the right-hand side we have
\be\nonumber
 \epsilon^{\mu\nu\rho\sigma}\left(\frac{1}{2t} \bar{\Omega}_{\nu}{}^\alpha \partial_{\mu} \partial_\alpha \phi  - \frac{1}{2t^2} (dt-\im dx)_{\mu} \bar{\Omega}_{\nu}{}^\alpha \partial_\alpha \phi \right)
 \left(\frac{1}{2t} \bar{\Omega}_{\sigma}{}^\beta \partial_{\rho} \partial_\beta \phi -\frac{1}{2t^2} (dt-\im dx)_{\rho} \bar{\Omega}_{\sigma}{}^\beta \partial_\beta \phi \right) = \\ \nonumber
   \frac{1}{4t^2}  \epsilon^{\mu\nu\rho\sigma} \bar{\Omega}_{\nu}{}^\alpha \partial_{\mu} \partial_\alpha \phi  \bar{\Omega}_{\sigma}{}^\beta \partial_{\rho} \partial_\beta \phi 
   - \frac{1}{2t^3} \epsilon^{\mu\nu\rho\sigma}\bar{\Omega}_{\nu}{}^\alpha \partial_{\mu} \partial_\alpha \phi (dt-\im dx)_{\rho} \bar{\Omega}_{\sigma}{}^\beta \partial_\beta \phi .
 \ee
 We again use (\ref{epsilon-omega}), which we write as
 \be
\epsilon^{\mu\nu\rho\sigma} \bar{\Omega}_{\nu}{}^{\alpha} = g^{\rho\alpha} \bar{\Omega}^{\mu\sigma} + g^{\mu\alpha} \bar{\Omega}^{\sigma\rho}+g^{\sigma\alpha} \bar{\Omega}^{\rho\mu}.
\ee
Then the right-hand side becomes
\be
 -\frac{1}{4t^2}   \bar{\Omega}^{\mu\rho} \partial_{\mu} \partial_\alpha \phi  \bar{\Omega}^{\alpha\beta} \partial_{\rho} \partial_\beta \phi
  + \frac{1}{2t^3} \bar{\Omega}^{\mu\rho} \partial_{\mu} \partial_\alpha \phi (dt-\im dx)_{\rho} \bar{\Omega}^{\alpha\beta} \partial_\beta \phi .
   \ee
   Using
   \be
   \bar{\Omega}^{\mu\rho}(dt-\im dx)_{\rho}  = - 4t^2 \left(\frac{\partial}{\partial z}\right)^\mu,
   \ee
  we get the following equation:
  \be
    \frac{1}{t} \Box\phi=  \frac{1}{8t^4}   \bar{\Omega}^{\mu\rho} \partial_{\mu} \partial_\alpha \phi  \bar{\Omega}^{\alpha\beta} \partial_{\rho} \partial_\beta \phi
   + \frac{1}{t^3}  \partial_{z} \partial_\alpha \phi  \bar{\Omega}^{\alpha\beta} \partial_\beta \phi .
   \ee
   This is (\ref{arthur-formula}) written in a covariant form, which is our main new result. It can also be rewritten more compactly as
    \be\label{sdgnew}
    \frac{1}{t} \Box\phi=  \frac{1}{8t^4}   \bar{\Omega}^{\mu\rho} \partial_{\mu} \partial_\alpha \phi  \, \bar{\Omega}^{\alpha\beta} (\partial_{\rho} - \frac{2}{t} (dt-\im dx)_{\rho} )  
  \partial_\beta \phi.
  \ee

\section{Conclusion}

In this paper we provide a concise new derivation of the constant curvature background version (\ref{arthur-formula}) of the Plebanski second heavenly equation (\ref{sd-eqs-flat}). We also rewrite this equation in the new form \eqref{sdgnew}, which is considerably more compact than (\ref{arthur-formula}). This exhibits the clear parallel with the flat version of this equation in the form (\ref{sd-eqs-flat}). The key ingredients were the pure connection formalism and a covariant ansatz for the connection in \eqref{a-Omega} which was inspired by a similar ansatz for the gauge field in SDYM in \eqref{param-A-YM}. The key ingredient of the ansatz is a complex 2-form $\bar{\Omega}$ which encodes the choice of null coordinates, or equivalently the choice of complex structure. This in turn suggests a new interpretation for the kinematic aglebra as of SDYM as the Lie algebra of $(0,1)$ vector fields on a $\R^4$ endowed with a complex structure.

This work opens a number of avenues for future research. Perhaps the most immediate goal would be to use the formalism developed in this paper to compute boundary correlators of SDGR in a constant curvature space and compare to the results recently obtained in \cite{Chowdhury:2024dcy}. This reference computes correlators of the scalar potential $\phi$. The formalism we described presents an alternative. It is not difficult to work with the covariant SDGR Lagrangian (\ref{SDGR-Lagr}) directly, by appropriately gauge-fixing the kinetic term to obtain the propagator. A way to do this gauge-fixing is described in \cite{Krasnov:2016emc}. The ansatz \eqref{a-Omega} should then be interpreted as one to be imposed on the on-shell external legs of the Feynman diagrams to be computed, where an appropriate choice of the potential $\phi$ would parametrise the graviton polarisations. In \cite{Chowdhury:2024dcy} it was also observed that different choices of lightcone gauge have the potential to simplify the results and that perturbation theory may be simplest if the lightcone direction lies along the boundary of AdS$_4$. The covariant formulation presented in this paper may therefore be very convenient because it allows for a general choice of the null direction. Moreover the complex 2-form $\bar{\Omega}$ can be readily expressed in terms of a reference spinor making the mapping to spinor-helicity variables more straightforward. 

In flat space, all-multiplicity formulae for MHV amplitudes in full YM and GR can be derived \cite{Krasnov:2016emc} using the technology of Berends-Giele currents \cite{Berends:1987me}, applied to the self-dual sectors. More recently, this was also carried out using the pertubiner expansion \cite{Miller:2024oza}. Berends-Giele recursion was also recently generalised to full YM and GR in (A)dS in \cite{Armstrong:2022mfr}, see also \cite{Albrychiewicz:2021ndv}. It would therefore be very interesting to attempt to use Berends-Giele recursion to compute boundary correlators of SDYM and SDG in (A)dS$_4$ in the hope that the integrability of these theories allows one to derive explicit formulae for any multiplicity. Thus, one could hope for a $\Lambda\not=0$ version of the Parke-Taylor formula for gluons \cite{Parke:1986gb} or the Hodges formula for gravitons \cite{Hodges:2012ym}. We hope to explore this in future work.

\section*{Acknowledgements}
AL and KK are grateful to the organisers of the workshop “Twistors and Higher Spins” which took place in July 2024 at University of Mons, where this work was initiated. AL is supported by an STFC Consolidated Grant ST/T000708/1.

\end{document}